\documentclass[fleqn,12pt,twoside]{article}
\usepackage{espcrc1}
\readRCS
$Id: espcrc1.tex,v 1.2 2004/02/24 11:22:11 spepping Exp $
\usepackage{graphicx}
\usepackage[figuresright]{rotating}

\newcommand{\AmS}{{\protect\the\textfont2
  A\kern-.1667em\lower.5ex\hbox{M}\kern-.125emS}}

\title{Transverse Spectra of Hadrons in Central $AA$ Collisions at RHIC and LHC from pQCD+Saturation+Hydrodynamics and from pQCD+Energy Losses}

\author{K. J. Eskola\address[JYFL]{Department of Physics,
	P.B. 35, FIN-40014 University of Jyv\"askyl\"a, Finland}\address[HIP]{Helsinki Institute of Physics, 
	P.B. 64, FIN-00014 University of Helsinki, Finland},
	H. Honkanen\addressmark[JYFL]\addressmark[HIP]\address{Department of Physics,
	University of Virginia, P.O.B. 400714, Charlottesville, VA 22904-4714, USA},
	\underline{H. Niemi}\addressmark[JYFL]\addressmark[HIP],
	P. V. Ruuskanen\addressmark[JYFL]\addressmark[HIP],
	and
	S. S. R\"as\"anen\addressmark[JYFL].
}

\begin{document}

% typeset front matter
 \voffset=-1cm
\maketitle

\begin{abstract}

We study the transverse spectra of hadrons in nearly central $AA$
collisions at RHIC and LHC in a broad transverse momentum range~\cite{Eskola:2005ue}.
Low-$p_T$ spectra are calculated by using boost-invariant
hydrodynamics with initial energy and net-baryon densities from the EKRT~\cite{EKRT}
pQCD+saturation model.  High-$p_T$ spectra are obtained from pQCD
jet calculation~\cite{EH03} including the energy loss of the parton~\cite{EHSW04} in the
matter prior to its fragmentation to final hadrons.

\end{abstract}

\section{Introduction}

Transverse momentum spectra of hadrons in ultrarelativistic nuclear
collisions provide valuable information of particle production
mechanism in the collisions as well as dynamics and properties of
the produced QCD matter. Low-$p_T$ features of the single particle
spectra are well described by hydrodynamical models and the data is
consistent with an ideal fluid behaviour of the matter.  The
observed azimuthal asymmetry in non-central Au+Au collisions at RHIC
has been argued to result from strong collective motion and
early thermalization of produced partonic matter. The suppression
observed in the high-$p_T$ tail of the spectra relative to the p+p and
d+Au collisions is understood as the energy loss of high-$p_T$
partons into the thermalized partonic matter.

Our approach is to calculate initial particle production using the EKRT
model which is based on the idea that the low-$p_T$ particle
production is controlled by saturation among the final state
gluons in contrast to the initial state saturation models, where
saturation is the property of colliding nuclei.
% From the model we calculate initial energy and net-baryon
% densities in nuclear collisions.
The EKRT model provides a closed framework to calculate initial transverse
energy and net-baryon number in midrapidity in nuclear collisions
with sufficiently large $\sqrt{s}$ and A. Final low-$p_T$ hadron
spectra are calculated by using hydrodynamics with initial densities
from the EKRT model. A good agreement with the measured data in central
Au+Au collisions at RHIC is obtained~\cite{Eskola:2005ue,Eskola:2002wx}.
% The full model is applied to central Au+Au
% collisions at RHIC and it is shown to be in a good agreement with
% the measured data~\cite{Eskola:2005ue}.  
We further predict the hadron spectra in central
Pb+Pb collisions at the LHC~\cite{Eskola:2005ue}.

High-$p_T$ part of the spectra is calculated using factorized pQCD
parton model for high-$p_T$ parton production and taking into
account the parton energy loss before fragmentation.
% We use leading order pQCD cross sections with K-factors, that are
% fixed from $p+p(\bar{p})$ data.  Transport coefficient for energy
% loss is fixed from RHIC $\sqrt{s_{NN}}$=200 GeV Au+Au data.

We compare these two models with the RHIC data and determine the 
regions where each component is dominant. In particular, we find 
that the low-$p_T$ hydrodynamical spectrum dominates over the fragmentation
spectrum in a much wider $p_T$ region at the LHC than at RHIC.  We
discuss the independence of the two components and conclude that
they are most likely almost independent even in the cross-over
region.

\section{Models}

\subsection{pQCD + saturation + hydrodynamics} 
The EKRT model~\cite{EKRT} estimates the final state saturation scale using 
the following geometric criterion: Saturation becomes important 
when the produced gluons fill the whole available transverse 
area of the colliding nuclei. For central collisions this can 
be written as
\begin{equation}
N_{AA}(q_0, \sqrt{s}, A) \frac{\pi}{q_0^2} = \pi R_A^2,
\end{equation}
where $N_{AA}$ is the number of gluons above a transverse momentum cut-off, $q_T >
q_0$ in the rapidity interval $|y|\leq 0.5$, and $R_A$ is 
the nuclear radius. This condition provides the saturation scale 
$p_{sat}=q_0$.  If $p_{sat} \gg \Lambda_{QCD}$,
pQCD can be used to estimate the number of produced partons and the
amount of transverse energy at midrapidity.  This approach gives
also the net-baryon number at midrapidity.  If we assume that the
produced matter thermalizes immediately after production at
$\tau_0=1/p_{sat}$, we obtain the initial energy and net-baryon 
density at $\tau_0$ for hydrodynamical evolution.

We use ideal fluid hydrodynamics with boost-invariance and azimuthal 
symmetry. In the bag-model equation of state the hadron gas phase 
consists of all hadronic states with $m<2$ GeV and the QGP phase 
of massless gluons and three flavors of quarks. Critical temperature 
is chosen to be $165$ MeV.
After the hydrodynamic expansion and cooling of the matter, the
Cooper-Frye decoupling prescription is applied for the calculation
of the low-$p_T$ spectra.  Below we show the sensitivity of the
decoupling procedure on the decoupling condition. 

\subsection{pQCD + fragmentation + energy loss}

High-$p_T$ spectra are calculated using nuclear parton distributions,
pQCD parton cross sections, fragmentation functions and quenching
weights for energy losses.  We use leading order perturbative cross
sections with K-factors fixed from $p+p(\bar{p})$ data~\cite{EH03}.  
The K-factors are extrapolated to the LHC energy by using different 
parametrizations to estimate the uncertainties in the extrapolation.
% Energy losses are included using quenching weight.
The magnitude of the energy losses can be expressed with one
effective transport coefficient, which is fixed from the RHIC Au+Au data
at $\sqrt{s_{NN}} = 200$ GeV~\cite{EHSW04}.  Since the transport coefficient is
proportional to energy density, fixing it in one collision predicts
it for other collisions. There is quite a large uncertainty
associated with the eikonal approximation used in the energy loss
calculation; arbitrarily large energy losses are allowed whereas 
the energy of real jets is limited. There are different ways to 
deal with this~\cite{EHSW04}, shown here as an uncertainty in our pQCD
fragmentation + energy loss results.

% \vspace{-1cm}
\section{Results}

\begin{figure}[!ht]
	\vspace{-1.9cm}
	\hspace{-0.8cm}
\begin{minipage}[t]{75mm}
\includegraphics[width=90mm]{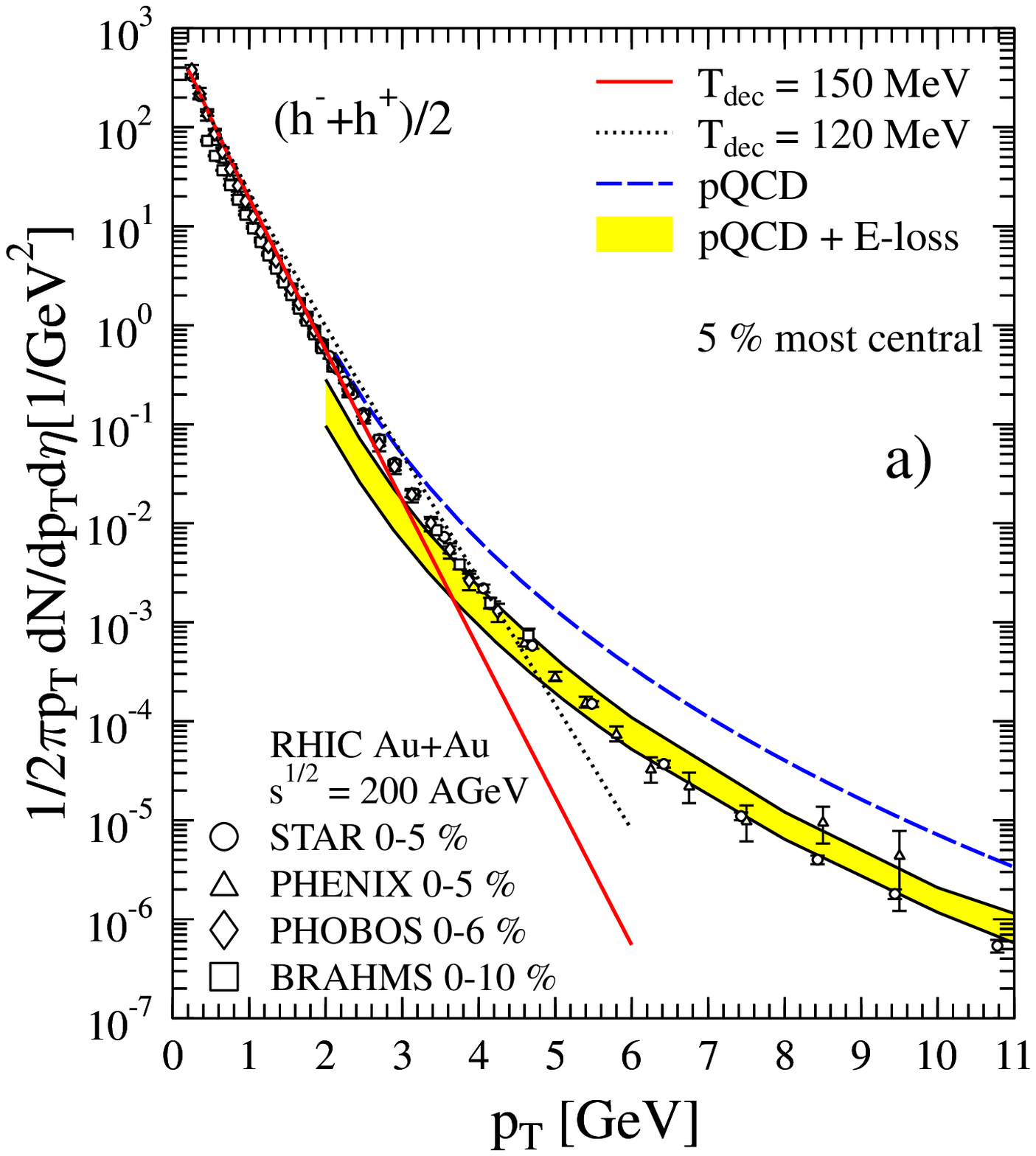}
\end{minipage}
 \hspace{\fill}
\begin{minipage}[t]{75mm}
 	\hspace{-0.6cm}
\includegraphics[width=90mm]{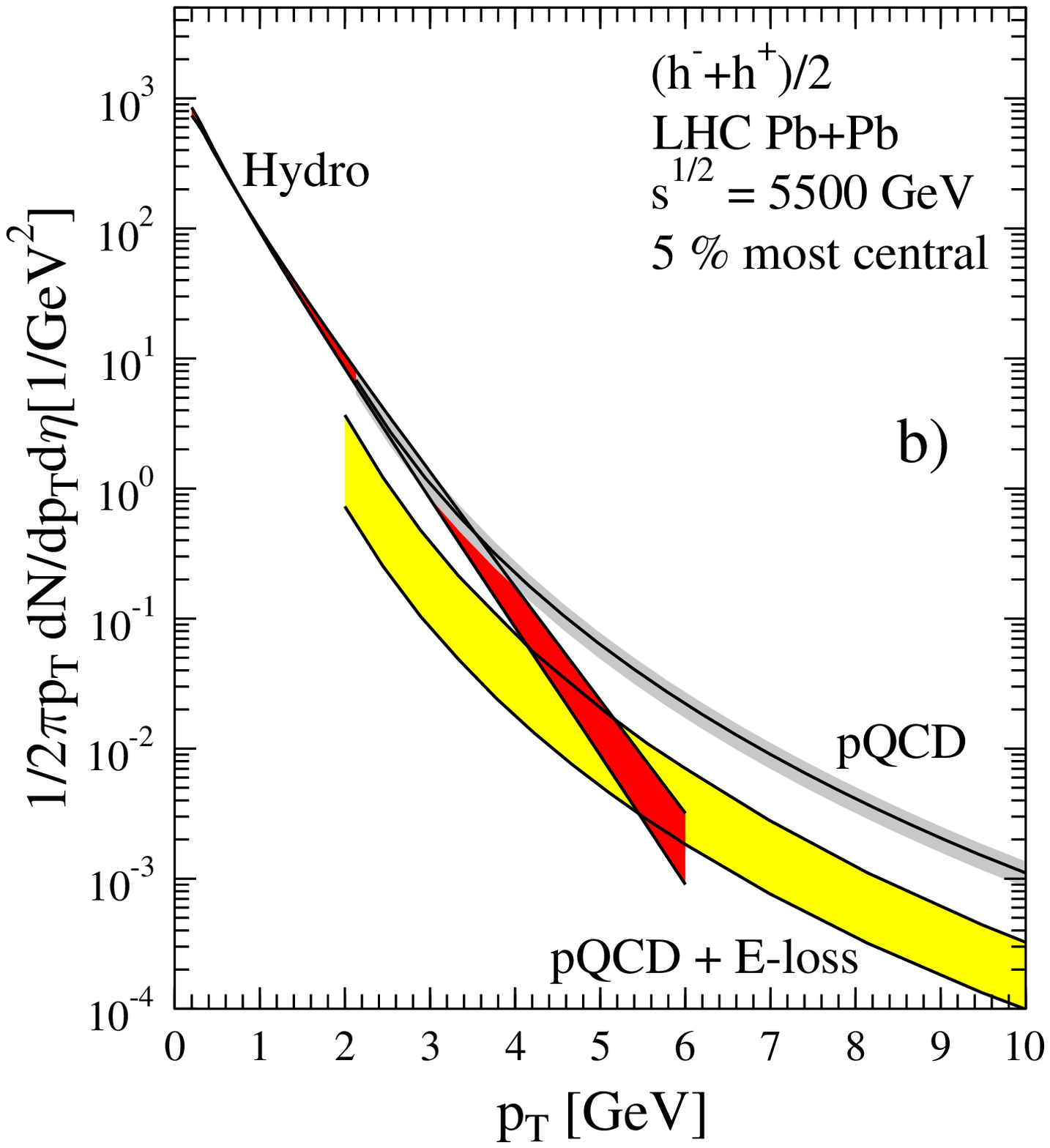}	
\end{minipage}
	\vspace{-1.5cm}
\caption{\protect\small
{\bf a)} Transverse momentum spectra of charged hadrons at $y=0$
in 5 \% most central Au+Au collisions at $\sqrt{s_{NN}}=200$~GeV.
The solid and dotted lines shows our hydrodynamic results with $T_{dec}=150$~MeV and
$T_{dec}=120$~MeV respectively. The shaded band shows the pQCD fragmentation+energy
loss spectrum and dashed line pQCD fragmentation without energy loss.
The data is from Refs.~\cite{Adams:2003kv,Adler:2003au,Back:2003qr,Arsene:2003yk}.
{\bf b)} As Fig.~\ref{fig:charged200}a but for the 5 \% most central Pb+Pb collision
at $\sqrt{s_{NN}}=5500$~GeV.
}
\label{fig:charged200}
\end{figure}

In Fig~\ref{fig:charged200}a we show our results for unidentified
charged hadrons at midrapidity for 5 \% most central Au+Au
collisions at $\sqrt{s_{NN}} = 200$ GeV.  As mentioned above we show our hydro results
with two different decoupling temperatures to show the sensitivity
on the decoupling condition.  From the figure we observe that both
the normalization and slope of the spectra are well reproduced with
a single decoupling temperature $T_{dec} = 150$ MeV. The spectra 
from fragmentation calculation with and without energy loss are plotted
in the same figure. It is seen clearly that the measured
spectra at high $p_T$ cannot be explained without energy loss.  The
transport coefficient that determines the energy loss is fixed from
this data so that in this case the agreement is obtained by
construction. The band in the energy loss results shows the uncertainty
of eikonal approximation. For identified particles the spectra can be
found from~\cite{Eskola:2005ue}.

Fig~\ref{fig:charged200}b shows our prediction for the 5 \% most central
Pb+Pb collisions at the LHC.  The hydrodynamical results are again
shown as a band corresponding to the decoupling temperatures between 
$120$ and $150$ MeV. The band for the pQCD fragmentation results 
without energy loss is from the uncertainty in the extrapolation of 
the K-factors to the LHC energy. The large uncertainty in the energy loss 
calculation comes, as before, from the eikonal approximation.

We see in Fig~\ref{fig:charged200}a that at RHIC the hydrodynamical
and pQCD + energy loss spectra cross at $p_T \sim 3\dots4$ GeV, and
that the data starts to deviate from the hydro spectrum in the same $p_T$
region.  At the LHC, the crossing
region is moved to $p_T \sim 5 \dots6$ GeV, suggesting a wider $p_T$
region of applicability of hydrodynamics at the LHC than at RHIC.
It is also interesting to note that in the cross-over region the
fragmentation and the hydrodynamical components are most likely
almost independent:  At RHIC 95 \% of the thermalized matter comes
from mini-jets with partonic transverse momenta $p_{sat} < q_T <
3.6$ GeV and the higher-$q_T$ partons contribute to the normalizations and
the slopes of the hydrodynamic spectra only slightly.  On the other
hand even without the energy loss the pQCD pions are dominantly from
partons with $q_T \sim 1.7 p_T$, which is $5.1$ GeV for $p_T \sim 3$
GeV pions.  With energy loss included they originate from even higher
$q_T$.  Thus the partonic origin of fragmentation pions and pions
from thermalized matter is quite different suggesting that the two
contributions are quite independent even in the cross-over region
and can be added without serious double counting.
The same argument holds also at the LHC. Crossing between hydro and
fragmentation spectra depends also on particle species as studied at
RHIC energies in ref.~\cite{Hirano:2003pw}.

\section{Conclusions}

We have calculated low-$p_T$ spectra of hadrons for central Au+Au
collisions at RHIC and Pb+Pb collisions at the LHC using the EKRT model
to calculate the initial parton production and hydrodynamics to
calculate the expansion of produced matter.  High-$p_T$ hadron
spectra are calculated by assuming that the high-$q_T$ partons do
not thermalize but fragment to hadrons after loosing energy when
traversing the thermal matter.  Our model is shown to be in a good
agreement with the measured data in central collisions at RHIC. 
We have provided predictions for central Pb+Pb collisions at the LHC, including also
the net baryon number, see~\cite{Eskola:2005ue}.
The origins of the thermal and the pQCD fragmentation spectra 
are discussed and it is argued that even in the cross-over region 
where the two components are comparable, they are essentially independent.

%%%%%%%%%%%%%%%%%%%%% BEGIN FIGURE %%%%%%%%%%%%%%%%%%%%%%%%%%%%%%%%
%%%%%%%%%%%%%%%%%%%%% END FIGURE %%%%%%%%%%%%%%%%%%%%%%%%%%%%%%%%


\vspace{-1cm}

\begin{thebibliography}{9}

%\cite{Eskola:2005ue}
\bibitem{Eskola:2005ue}
K.~J.~Eskola, H.~Honkanen, H.~Niemi, P.~V.~Ruuskanen and S.~S.~R\"as\"anen,
%``RHIC-tested predictions for low-p(T) and high-p(T) hadron spectra in nearly
%central Pb + Pb collisions at the LHC,''
arXiv:hep-ph/0506049.
%%CITATION = HEP-PH 0506049;%%

%\cite{EKRT}
\bibitem{EKRT}
K.~J.~Eskola, K.~Kajantie, P.~V.~Ruuskanen and K.~Tuominen,
%``Scaling of transverse energies and multiplicities with atomic number  and
%  energy in ultrarelativistic nuclear collisions,''
Nucl.\ Phys.\ B {\bf 570} (2000) 379
[arXiv:hep-ph/9909456].
%%CITATION = HEP-PH 9909456;%%

\bibitem{EH03}
%\bibitem{Eskola:2002kv}
K.~J.~Eskola and H.~Honkanen,
%``A perturbative QCD analysis of charged-particle distributions in  hadronic
%and nuclear collisions,''
Nucl.\ Phys.\ A {\bf 713} (2003) 167
[arXiv:hep-ph/0205048].
%%CITATION = HEP-PH 0205048;%%

\bibitem{EHSW04}
%\bibitem{Eskola:2004cr}
K.~J.~Eskola, H.~Honkanen, C.~A.~Salgado and U.~A.~Wiedemann,
%``The fragility of high-p(T) hadron spectra as a hard probe,''
Nucl.\ Phys.\ A {\bf 747} (2005) 511
[arXiv:hep-ph/0406319].
%%CITATION = HEP-PH 0406319;%%

%\cite{Eskola:2002wx}
\bibitem{Eskola:2002wx}
K.~J.~Eskola, H.~Niemi, P.~V.~Ruuskanen and S.~S.~R\"as\"anen,
%``Dependence of hadron spectra on decoupling temperature and resonance
%contributions,''
Phys.\ Lett.\ B {\bf 566} (2003) 187
[arXiv:hep-ph/0206230].
%%CITATION = HEP-PH 0206230;%%

%\cite{Hirano:2003pw}
\bibitem{Hirano:2003pw}
T.~Hirano and Y.~Nara,
%``Interplay between soft and hard hadronic components for identified  hadrons
%in relativistic heavy ion collisions at RHIC,''
Phys.\ Rev.\ C {\bf 69} (2004) 034908
[arXiv:nucl-th/0307015].
%%CITATION = NUCL-TH 0307015;%%

%\bibitem{HIPT_STAR}
\bibitem{Adams:2003kv}
J.~Adams {\it et al.}  [STAR Collaboration],
%``Transverse momentum and collision energy dependence of high p(T) hadron
%suppression in Au + Au collisions at ultrarelativistic energies,''
Phys.\ Rev.\ Lett.\  {\bf 91} (2003) 172302.
% [arXiv:nucl-ex/0305015].
%%CITATION = NUCL-EX 0305015;%%

\bibitem{Adler:2003au}
S.~S.~Adler {\it et al.}  [PHENIX Collaboration],
%``High-p(T) charged hadron suppression in Au + Au collisions at
%  s(NN)**(1/2) = 200-GeV,''
Phys.\ Rev.\ C {\bf 69} (2004) 034910.
% [arXiv:nucl-ex/0308006].
%%CITATION = NUCL-EX 0308006;%%

%\bibitem{HIPT_PHOBOS}
\bibitem{Back:2003qr}
B.~B.~Back {\it et al.}  [PHOBOS Collaboration],
%``Charged hadron transverse momentum distributions in Au + Au collisions  at
%s(NN)**(1/2) = 200-GeV,''
Phys.\ Lett.\ B {\bf 578} (2004) 297.
% [arXiv:nucl-ex/0302015].
%%CITATION = NUCL-EX 0302015;%%

%\bibitem{HIPT_BRAHMS}
\bibitem{Arsene:2003yk}
I.~Arsene {\it et al.}  [BRAHMS Collaboration],
%``Transverse momentum spectra in Au + Au and d + Au collisions at
%  s(NN)**(1/2) = 200-GeV and the pseudorapidity dependence of high
%  p(T)  suppression,''
Phys.\ Rev.\ Lett.\  {\bf 91} (2003) 072305.
% [arXiv:nucl-ex/0307003].
%%CITATION = NUCL-EX 0307003;%%


\end{thebibliography}
\end{document}